# All-angle negative refraction of highly squeezed plasmon and phonon polaritons in graphene-boron nitride heterostructures

Xiao Lin[a,b,1,2], Yi Yang[c,1], Nicholas Rivera[b], Josué J. López[c], Yichen Shen[b], Ido Kaminer[b,3], Hongsheng Chen[a,d,3], Baile Zhang[e,f,3], John D. Joannopoulos[b,3], and Marin Soljačić[b]

[a]*State Key Laboratory of Modern Optical Instrumentation, Zhejiang University, Hangzhou 310027, China.*

[b]*Department of Physics, Massachusetts Institute of Technology, Cambridge, MA 02139, USA.*

[c]*Department of Electrical Engineering and Computer Science, Massachusetts Institute of Technology, Cambridge, MA 02139, USA.*

[d]*The Electromagnetics Academy at Zhejiang University, State Key Laboratory of Modern Optical Instrumentation, Zhejiang University, Hangzhou 310027, China.*

[e]*Division of Physics and Applied Physics, School of Physical and Mathematical Sciences, Nanyang Technological University, Singapore 637371, Singapore.*

[f]*Centre for Disruptive Photonic Technologies, Nanyang Technological University, Singapore 637371, Singapore.*

**Abstract:** A fundamental building block for nanophotonics is the ability to achieve negative refraction of polaritons, because this could enable the demonstration of many unique nanoscale applications such as deep-subwavelength imaging, superlens, and novel guiding. However, to achieve negative refraction of highly squeezed polaritons, such as plasmon polaritons in graphene and phonon polaritons in boron nitride (BN) with their wavelengths squeezed by a factor over 100, requires the ability to flip the sign of their group velocity at will, which is challenging. Here we reveal that the strong coupling between plasmon and phonon polaritons in graphene-BN heterostructures can be used to flip the sign of the group velocity of the resulting hybrid (plasmon-phonon-polariton) modes. We predict all-angle negative refraction between plasmon and phonon polaritons, and even more surprisingly, between hybrid graphene plasmons, and between hybrid phonon polaritons. Graphene-BN heterostructures thus provide a versatile platform for the design of nano-metasurfaces and nano-imaging elements.

**Keywords:** negative refraction | plasmon polariton | phonon polariton | graphene-boron nitride heterostructure

Polaritons with high spatial confinement, such as plasmon polaritons in graphene [1-5] and phonon polaritons in a thin hexagonal boron nitride (BN) slab [6-15], enable control over the propagation of light at the extreme nanoscale, thanks to their in-plane polaritonic wavelength that can be squeezed by a factor over 100. Henceforth we use the term squeezing factor (or confinement factor) to define the ratio between the wavelength in free space and the in-plane polaritonic wavelength. The combination of tunability, low losses and ultra-confinement [1,2,8,10,11,15] makes them superior alternatives to conventional metal plasmons and highly appealing for nanophotonic applications [3-5,10-13,15]. Their extreme spatial confinement, however, also limits our ability to tailor their dispersion relations.

Unlike the case of 2D plasmons, the coupling between metal plasmons in a metal-dielectric-metal structure dramatically changes their dispersion relation and can even flip the sign of their group velocities [16,17]. This has led to exciting applications by tailoring the in-plane plasmonic refraction, giving flexibility in controlling the energy flow of light. Specifically, by flipping the sign of the group velocity of metal plasmons, plasmonic negative refraction has been predicted [16] and demonstrated [17]. The negative refraction has also been extensively explored in metamaterials, metasurfaces and photonic crystals [18-26], but they become experimentally very challenging to realize when dealing with polaritons with high squeezing factors. In contrast to metal plasmons, the group velocity of graphene plasmons [2,11,27] and all other 2D plasmons [28-32] is always positive, including that in graphene-based multilayer structures [33]. This has made the in-plane negative refraction for highly squeezed 2D plasmon polaritons seem impossible to achieve.

Contrary to 2D plasmons, the group velocity of phonon polaritons in a BN slab is negative (positive) in BN's first (second) reststrahlen band [8,10,11], as was recently observed experimentally [9,12]. Yet, like 2D plasmons, the control of the group velocities of these phonon polaritons remains difficult due to their high spatial confinement [6,7,9,12,13]. Moreover, although recent works have demonstrated that the dispersion relations of graphene plasmons and BN's phonon polaritons in a graphene-BN heterostructure can be modified through plasmon-phonon-polariton hybridization [6,10,11,14], the ability to flip the sign of the group velocity of these strongly squeezed hybrid polaritons has not been observed nor predicted so far.

In this work, we predict that utilizing the strong coupling between graphene plasmon and BN's phonon polaritons in graphene-BN heterostructures provides an unprecedented control over their dispersion relations and can even flip the signs of their group velocities within BN's first reststrahlen band. Using realistic material losses in graphene-BN heterostructures, we demonstrate theoretically all-angle in-plane polaritonic negative refraction between graphene plasmon and BN's phonon polariton. And even more surprisingly, we predict negative refraction between two kinds of graphene plasmons and between two kinds of BN's phonon polaritons, where the squeezing factors of these polaritons are all larger than 100. Unlike conventional all-angle negative refraction of metal plasmons that are restricted to work at a given fixed frequency [16], the working frequency of the all-angle polaritonic negative refraction demonstrated here can be flexibly tunable within BN's first reststrahlen band, due to the tunability of the chemical potential in graphene.

**Results**

To highlight the underlying physics, we begin by studying the in-plane negative refraction between the graphene plasmon and BN's phonon polariton within BN's first reststrahlen band (i.e. 22.3 to 24.6 THz), as shown in Fig. 1. In the heterostructure [Fig.1*A*], the left air-graphene-substrate region supports plasmon polaritons, and the right air-BN-substrate region supports phonon polaritons. These polaritons propagate in the x-y plane, and there is an interface along the y-direction. In this work, the random phase approximation (RPA) [2,27,34] is used to characterize graphene's surface conductivity. A conservative electron mobility of 10000 $cm^2V^{-1}s^{-1}$ is assumed in graphene. Both realistic losses in graphene and BN are considered. In order to maintain the relatively low propagation loss of these highly squeezed polaritons, we consider lossless dielectrics as substrates, for example germanium [35,36] with a relative permittivity of $\varepsilon_r = 16$ at the frequency of interest. We note that although SiO$_2$ and SiC [36] are popular substrates for graphene and BN devices, they are lossy within BN's first reststrahlen band. In addition, the choice of bulk BN as the substrate is also not beneficial because the mode of graphene plasmons will disappear within BN's first reststrahlen band [11], as can be seen in Fig.S1.

Figure 1*B* shows the dispersion of graphene plasmons and BN's phonon polaritons. For clarity, below we discuss the direction of the group velocity $\partial\omega/\partial q$ of these polaritons and/or the direction of their power flow with respect to the direction of their phase velocity $\omega/q$; when the sign of the group velocity is defined as positive, the group velocity and the phase velocity are in the same direction; when the sign of the group velocity is defined as negative, the group velocity and the phase velocity are in the opposite direction. The group velocity of graphene plasmons is positive while that of BN's phonon polaritons is negative in Fig.1*B*. The direction of the group velocity is determined by the direction of the total power flow, which can be quantitatively described (see supporting info for calculation details). To gain an intuitive understanding of the group velocity of these polaritons, one should analyze the direction of their total power flow. For graphene plasmons, the power flow is positive in the surrounding dielectrics and is negligible within graphene due to graphene's infinitesimal thickness, rendering the group velocity of graphene plasmons to be positive. For BN's phonon polaritons, since BN is a type-I hyperbolic material ($\varepsilon_x = \varepsilon_y > 0$, $\varepsilon_z < 0$) within the first reststrahlen band [7,12], the power flow is negative inside BN. Moreover, there is more power flow inside BN than that outside BN, rendering the group velocity of BN's phonon polaritons to be negative.

The in-plane negative refraction between graphene plasmon and BN's phonon polariton can happen because of their opposite signs of group velocities and approximately the same squeezing factors within BN's first reststrahlen band. We can choose a working frequency such that the wavevector of these polaritons are equal, such as the blue solid point (Re(q) = $94 \times 10^6$ m$^{-1}$ at 22.96 THz) in Fig.1*B*. Therefore, the effective phase index of these polaritons at this specific point are matched in magnitude. Equally as important, this phase-index matching further enables the in-plane negative refraction *at all angles of incidence* [16,22].

As shown in Fig. 1*C*, we numerically demonstrate the all-angle in-plane negative refraction between graphene plasmons and BN's phonon polaritons at 22.96 THz using the finite-element method (COMSOL). Graphene plasmons are excited by a dipole source in the left region and couple to BN's phonon polaritons at the interface. Since the signs of the group velocities of polaritons in the left and right regions are opposite, we get negative refraction in the plane where these polaritons propagate (i.e. the x-y plane). More generally, this indicates that the negative refraction happens in a plane parallel to the graphene plane, which is thus denoted as in-plane negative refraction. This in-plane negative refraction happens even when the configuration has different thicknesses on the two sides of the interface. Moreover, an image is formed in the right region, which computationally validates the all-angle negative refraction. In this system, the ratio $\left|\frac{Re(q)}{Im(q)}\right|$, as the dimensionless figure of merit for propagation damping [11], is equal to 15 for graphene plasmons and 23 for BN's phonon polaritons. This indicates relatively-low propagation losses for these polaritons [11], which facilitates future experimental verifications and applications. In Fig.1$C$, the squeezing factor $\frac{Re(q)}{\omega/c} = 195$ indicates that when compared with the wavelength in free space, the polaritonic wavelength is squeezed by a factor of 195. We note that the imaging mechanism here is not a perfect image recovery. This is because there is reflection at the interface due to the modal-profile mismatch [see $E_z$ profiles in Fig.1$A$] between plasmon and phonon polaritons, and the losses (i.e. $Im(q)$) in the left and right regions are different in Fig.1$C$. To suppress the reflection, one way is to maximize the mode overlap between the incident and transmitted mode profiles at the discontinuous interface [16]. Note that although the reflection at the interface can affect the transmission, the field intensity of polaritons in the region of transmission also largely depends on their propagation loss. Therefore, a complete optimization of the field intensity in the region of transmission should consider both reflection and the propagation loss of polaritons (see more discussion in supporting info). In addition to the situation considered above, we also consider a structure where graphene exists in both the left and the right regions (which may be easier to fabricate) and demonstrate the all-angle in-plane negative refraction between graphene plasmons and BN's phonon polaritons (Fig.S2).

While the negative refraction between the graphene plasmon and BN's phonon polariton in Fig.1 are quite expected because of their opposite group velocity signs in BN's first reststrahlen band, we also found two completely unexpected types of negative refractions occurring in these systems. We show how negative refraction can occur between two kinds of hybrid graphene plasmons (Fig.2) or between two kinds of hybrid phonon polaritons (Fig.3).

Below we show the strong coupling between graphene plasmons and BN's phonon polaritons can flip the signs of the group velocities of the hybrid plasmon-phonon-polaritons in graphene-BN heterostructures, as shown in Figs.2-3. There are two different types of polaritonic dispersion lines in graphene-BN heterostructures. For the first type, there is always a part of the dispersion line existing outside BN's first reststrahlen band, while the other part of the dispersion line can exist inside BN's first reststrahlen band, which is more plasmon-polariton-like; below, we thus refer to this type of mode as the type-I hybrid polariton. For the second type, the dispersion line only exists within BN's first reststrahlen band, which is more phonon-polariton-like; below, we refer to this type of mode as the type-II hybrid polariton.

Figure 2 demonstrates the all-angle in-plane negative refraction between the graphene plasmon and the type-I hybrid polariton. Since the type-I hybrid polariton is more plasmon-polariton-like, one can refer to this type of refraction as the negative refraction between two kinds of graphene plasmons. In Fig.2*A*, the left air-graphene-substrate region supports graphene plasmons, and the right air-graphene-BN-graphene-substrate region supports type-I hybrid plasmon-phonon-polaritons. Here we choose the lossless polycrystalline CVD diamond [36-38] with a relative permittivity of $\varepsilon_r = 5.6$ as the substrate, since the synthetic diamond is lossless at the frequency of interest and is also readily available [37-38].

Figure 2*B-E* shows the evolution of the dispersion of type-I hybrid polaritons in the right region at different chemical potentials. When the chemical potential of two graphene layers is low [Fig.2*B*], the group velocity of type-I hybrid polaritons is positive. When we increase the chemical potential [$\mu_c \geq 0.25$ eV in Fig.2*C-E*], part of the dispersion relation within BN's first reststrahlen band exhibits a negative group velocity. This twisting of the dispersion of type-I hybrid polaritons can again be intuitively understood via the power flow calculation. In graphene-BN heterostructures, the power flow is negative inside BN and more power can flow inside BN than outside BN. For example, for the blue crossing point between the dispersion lines of the graphene plasmon and the type-I hybrid polariton in Fig.2*D*, the ratio between the power flow inside BN ($P_{inBN}$) and that outside BN ($P_{outBN}$) is $\frac{P_{inBN}}{P_{outBN}} = -11$. The domination of the power flow in BN renders the group velocity of type-I hybrid polaritons to be negative. As a comparison, Fig.2*F* shows the dispersion of graphene plasmons in a structure without BN, where the group velocity stays positive.

Importantly, due to the tunability of the chemical potential in graphene, the phase-index matched crossing points in Figs.1-2 are dynamically controllable within BN's first reststrahlen band. For example, when the chemical potential of graphene in the left region of Fig.1*A* changes from 0.1 to 0.6 eV, the frequency of the phase-index matched crossing point in Fig.1*B* varies from 22.77 to 23.58 THz; when the chemical potential of graphene in the left region of Fig.2*A* changes from 0.05 to 0.25 eV, the frequency of the phase-index matched crossing points in Fig.2*D* varies from 22.94 to 23.49 THz. Moreover, since the negative group velocity of the type-I hybrid polaritons can be realized in a wide range of chemical potential ($\mu_c \geq 0.25$ eV) in the right region of Fig.2*A*, we can also tune the frequency of the phase-index matched crossing point by changing the chemical potential in the right region of Fig.2*A*. The different chemical potentials of graphene in the left and right regions in Fig.2 may be achieved by following the methods proposed in Ref.[1]. These above advantages offer the possibility for tunable in-plane polaritonic negative refraction, which is desirable for many nanophotonic applications. As an example, we show that for the phase-index matched crossing point at 23.32 THz in Fig.2*D*, $Re(q) = 66.8 \times 10^6$ m$^{-1}$, the squeezing factor $\frac{Re(q)}{\omega/c} = 137$, and $|\frac{Re(q)}{Im(q)}|$ is equal to 14 and 20 for the graphene plasmon and the type-I hybrid polariton, respectively. We numerically verify the in-plane negative refraction between graphene plasmons and type-I hybrid polaritons at 23.32 THz in Fig.2*G*, where type-I hybrid polaritons are incident from the right region at an angle of 45° (there is no specific requirement on the incident angle; see more discussion of the setup in supporting info).

Figure 3 demonstrates the all-angle in-plane negative refraction between BN's phonon polaritons and type-II hybrid polaritons. Since the type-II hybrid polariton is more phonon-polariton-like, we refer to this type of refraction as the negative refraction between two kinds of phonon polaritons. In Fig.3*A*, the left air-BN-substrate region supports phonon polaritons, and the right air-BN-graphene-substrate region supports type-II hybrid polaritons. Figure 3*B-E* show the modification of the sign of the group velocity for type-II hybrid polaritons by changing the thickness of BN in the right region. When BN is thick [Fig.3*B*], the group velocity of type-II hybrid polaritons is negative. When we decrease the thickness of BN, part of the dispersion line exhibits a positive group velocity [Fig.3*C-E*]. We use the power flow calculation to explain this dispersion twisting. The power flow is positive outside BN, and for small enough BN thicknesses, more power can flow outside BN than that inside BN, flipping the group velocity to be positive. Here we use the blue crossing point between the dispersion lines of BN's phonon polariton and the type-II hybrid polariton in Fig.3*D* as an example. The ratio between the power flow inside and outside BN is $\frac{P_{inBN}}{P_{outBN}} = -0.4$. Such a power flow distribution renders the group velocity of type-II hybrid polaritons to be positive. Since the BN slab behaves like a waveguide, we can also see that the number of the type-II hybrid polariton modes decreases when the thickness of BN decreases. As a comparison, Fig.3*F* shows the dispersion of BN's phonon polaritons in a structure without graphene, where the group velocity remains negative.

At the phase-index matched crossing point at 23.94 THz in Fig.3*D*, we have $Re(q) = 77.4 \times 10^6$ m$^{-1}$ and the squeezing factor $\frac{Re(q)}{\omega/c} = 154$; $|\frac{Re(q)}{Im(q)}|$ is equal to 21 and 9 for BN's phonon polariton and the type-II hybrid polariton, respectively. We numerically verify the in-plane negative refraction between BN's phonon polaritons and type-II hybrid polaritons in Fig.3*G*, where type-II hybrid polaritons are incident from the right region at an angle of 30º (the incident angle is arbitrarily chosen; see more discussion of the setup in supporting info).

**Discussion**

It is worthy to note that the all-angle negative refraction shown in Figs.2-3 is not suitable for dipole source imaging. This is because in the right region of the heterostructure, there are multiple hybrid plasmon-phonon-polariton modes with different wavevectors existing at the same working frequency [see dispersions in Fig.2*D* and Fig.3*D*]. When a dipole source is used, there will be unavoidable excitation of other hybrid polariton modes in the right region, which will blur the formation of an image.

Other types of negative refractions have been described in previous works, including the negative refraction of electrons [39-41]. Other studies have shown plasmonic [42-44] and optical [45,46] negative refraction in 3D bulk media made of graphene-based periodic structures. Negative refraction of light propagating through a single layer of graphene was first observed in Ref.[47]. However, most of these electromagnetic refractions occurred out of the graphene plane and not in-plane, therefore without taking advantage of the high spatial confinement and strong coupling of 2D polariton modes that are critical in this work. One viable way to experimentally verify the in-plane negative refraction of these highly squeezed polaritons is to apply the method of the direct geometric visualization of negative refraction [17].

In conclusion, we reveal a viable way to flip the sign of group velocities of hybrid plasmon-phonon-polaritons in graphene-BN heterostructures, by using the strong coupling between graphene plasmons and BN's phonon polaritons. This enables the flexible control of the in-plane refraction of highly squeezed polaritons, which is of fundamental importance for the manipulation of light at the nanoscale. Our full-wave simulations verify that all-angle in-plane negative refraction can be realized between graphene plasmons, BN's phonon polaritons and their hybrid polaritons in graphene-BN heterostructures. Moreover, the working frequency of these all-angle negative refraction can be flexibly tunable within BN's first reststrahlen band through changing the chemical potential of graphene. Due to the combined advantages of tunability, low loss, and superior spatial confinement provided by these polaritons on the platform of graphene-BN heterostructures, we expect many other novel actively-tunable polaritonic effects to be explored and utilized for the design of advanced polaritonic devices such as metasurfaces and superlenses.

**Materials and Methods**

The finite-element simulation is implemented via the frequency domain simulation in COMSOL. Graphene in the COMSOL simulation is modeled by a surface, where conditions for discontinuities in the electromagnetic fields are satisfied by taking into account the surface conductivity. This ensures high calculation accuracy, compared to the volumetric permittivity of an ultrathin graphene slab that is often considered [1]. The surface conductivity of the surface is calculated by the random phase approximation (RPA). The meshing resolution in the vicinity of graphene and BN is 1 nm in the z-direction. In Fig. 1, a z-polarized dipole source is placed in the left region and 20 nm above the structure to excite the plasmon polaritons. Fig. 1*C* is obtained at 20 nm below the BN-substrate interface. The phenomenon of negative refraction in Fig.1*C* is pronounced, regardless of the vertical position of the dipole source. In Figs. 2-3, a mode source is launched from the right side of the structure to excite the specific mode of interest. Figs. 2*G* and Fig.3*G* are obtained at 15 nm below and 2.5 nm above the BN-substrate interface, respectively.

**Footnotes**

[1]X.L. and Y.Y. contributed equally to this work.

[2]Present address: Division of Physics and Applied Physics, School of Physical and Mathematical Sciences, Nanyang Technological University, Singapore 637371, Singapore.

[3]To whom correspondence may be addressed. Email: kaminer@mit.edu, hansomchen@zju.edu.cn, blzhang@ntu.edu.sg, joannop@mit.edu.

Author contributions: X.L. and I.K. initiated the ide; X.L. performed the main analytical calculation; Y.Y. performed the main numerical calculation; X.L., Y.Y., N.R., J.J.L., Y.S., H.C., B.Z., J.D.J. and M.S. analyzed data, interpreted detailed results and contributed extensively to the writing of the manuscript; I.K., H.C., B.Z., J.D.J. and M.S. supervised the project.

The authors declare no competing financial interests.

This article contains supporting information online at www.pnas.org/lookup/suppl.

**ACKNOWLEDGMENTS.** This work was sponsored by the National Natural Science Foundation of China (Grants No. 61625502, 61574127 and 61601408), the ZJNSF (LY17F010008), the Top-Notch Young Talents Program of China, the Fundamental Research Funds for the Central Universities, the Innovation Joint Research Center for Cyber-Physical-Society System, Nanyang Technological University for NAP Start-Up Grant, the Singapore Ministry of Education (Grant No. MOE2015-T2-1-070, MOE2011-T3-1-005, and Tier 1 RG174/16 (S)), and the US Army Research Laboratory and the US Army Research Office through the Institute for Soldier Nanotechnologies (Contract No. W911NF-13-D-0001). M. Soljačić was supported in part (reading and analysis of the manuscript) by the MIT S3TEC Energy Research Frontier Center of the Department of Energy under Grant No. DESC0001299. I. Kaminer was partially supported by the Seventh Framework Programme of the European Research Council (FP7-Marie Curie IOF) under Grant No. 328853-MC-BSiCS. J. J. López is supported in part by a NSF Graduate Research Fellowship under award No. 1122374 and by a MRSEC Program of the National Science Foundation under award No. DMR-1419807. Y.Y. was partly supported by the MRSEC Program of the National Science Foundation under Grant No. DMR- 1419807.

[1] Vakil A, Engheta N (2011) Transformation optics using graphene. *Science* 332:1291-1294.

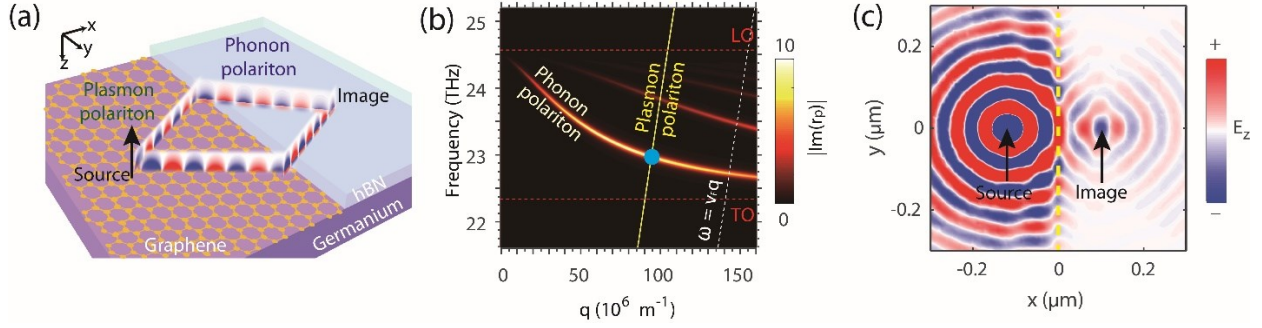

**Fig. 1.** In-plane negative refraction between plasmon polaritons in graphene and phonon polaritons in a BN slab. (*A*) Schematic structure, along with the $E_z$-field distribution of plasmon and phonon polaritons. The chemical potential of graphene is $\mu_c = 0.2$ eV and the thickness of BN is 20 nm. (*B*) Dispersion relation of phonon polaritons. It is an instructive way to visualize the dispersion of these polaritons via a false-color plot of $|\text{Im}(r_p)|$, where $r_p$ is the reflection coefficient of p-polarized waves for the right region in (*A*). This is because these polaritons are the singularity poles in the coefficient of $r_p$. The yellow line is the dispersion of plasmon polaritons for the left region in (*A*). The highlighted blue dot is the crossing point between the dispersion lines of plasmon and phonon polaritons. The red dashed lower and upper lines correspond to the transverse optical (TO) and longitudinal optical (LO) frequencies, respectively. The Fermi velocity is $v_F = c/300$. (*C*) $E_z$-field distribution excited by a dipole source at 22.96 THz. The yellow dashed line indicates the interface between left and right regions.

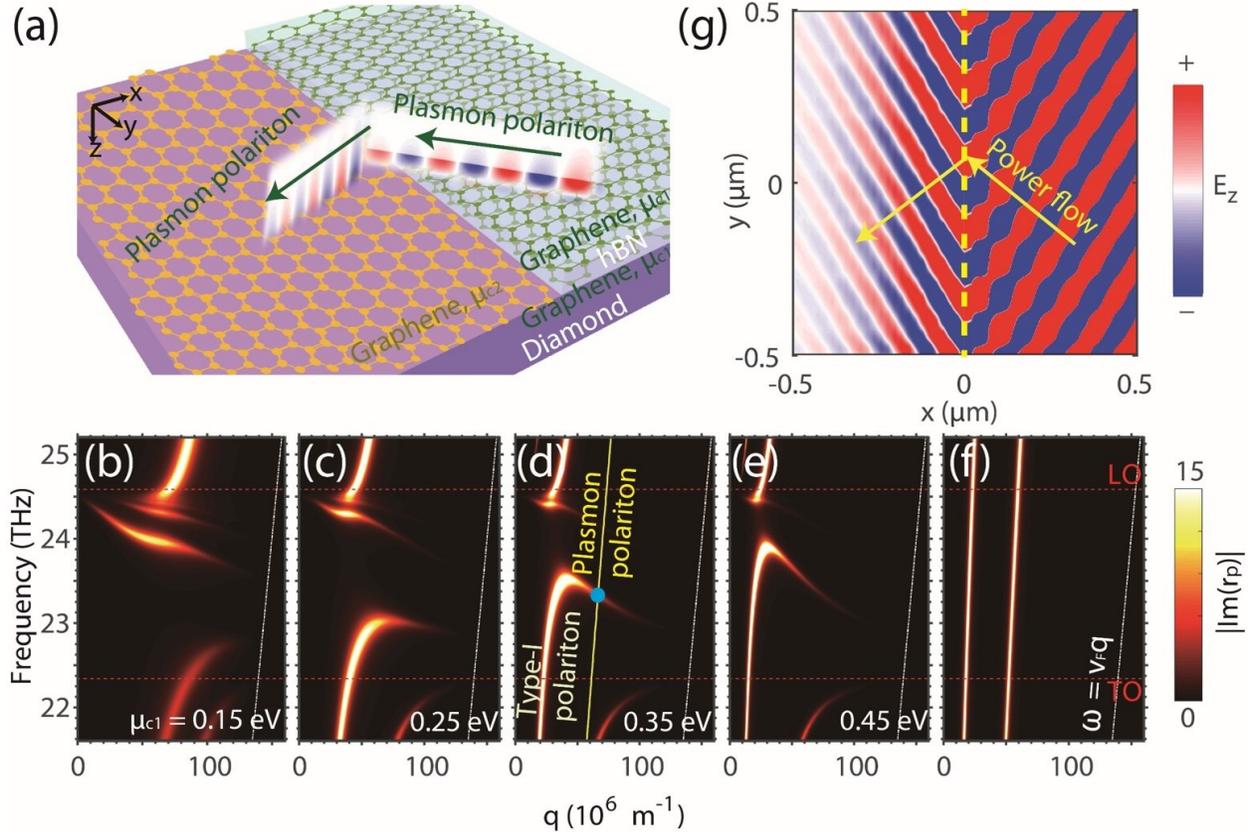

**Fig. 2.** In-plane negative refraction between graphene plasmons and plasmon-polariton-like type-I hybrid polaritons. (*A*) Schematic structure. The thickness of BN is 25 nm and the chemical potential of graphene in the left region is $\mu_{c2} = 0.15$ eV. (*B-E*) Evolution of the dispersion of type-I hybrid plasmon-phonon-polaritons for the right region in (*A*) under different values for the chemical potential, $\mu_{c1}$. The yellow line in (*D*) is the dispersion of graphene plasmons for the left region in (*A*). (*F*) Dispersion of air-graphene-substrate-graphene-substrate, where the structure is the same as that in (*D*) except for the replacement of BN with substrate. (*G*) $E_z$-field distribution of negative refraction. The incident and refracted polaritonic modes are highlighted by a blue solid point in (*D*) at 23.32 THz. The incident angle is 45° and the power flow direction is marked by the yellow arrows.

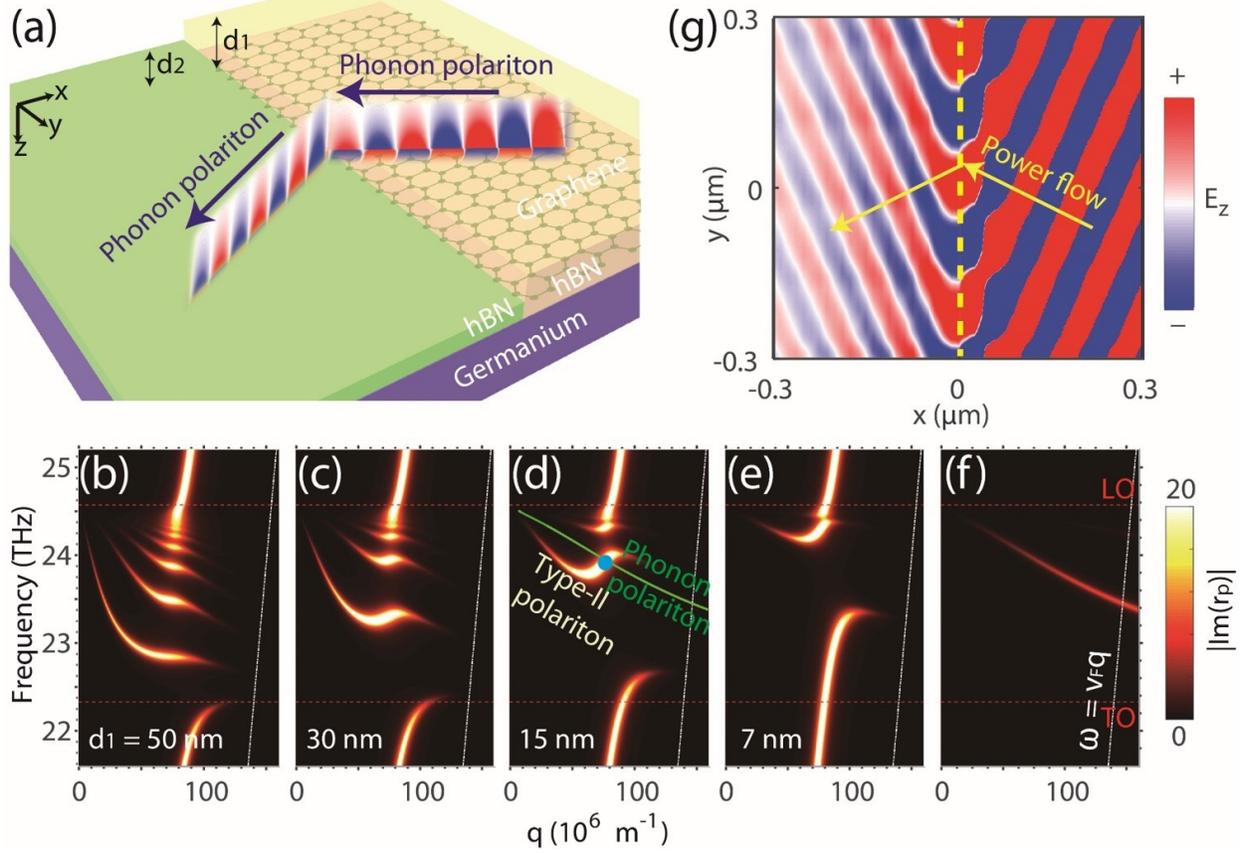

**Fig. 3.** In-plane negative refraction between BN's phonon polaritons and phonon-polariton-like type-II hybrid polaritons. (*A*) Schematic structure. The thickness of BN in the left region is $d_2 = 7$ nm and the chemical potential of graphene is $\mu_c = 0.3$ eV. (*B-E*) Evolution of the dispersion of type-II hybrid plasmon-phonon-polaritons for the right region in (*A*) under different thicknesses of the BN slab, $d_1$. The green line in (*D*) is the dispersion of BN's phonon polaritons (as shown in (*F*)). (*F*) Dispersion of air-BN-substrate, where the structure is the same as the left region in (*A*). (*G*) $E_z$-field distribution of negative refraction. The incident and refracted polaritonic modes are highlighted by a blue solid point in (*D*) at 23.94 THz. The incident angle is 30° and the power flow direction is marked by the yellow arrows.

16